\documentclass[usenatbib]{mn2e}
\newcommand{\bib}{\bibitem[\protect\citeauthoryear}
\newcommand{\pa}{\,\rlap{\raise 0.5ex\hbox{$\propto$}}{\lower 1.0ex\hbox{$\sim$}}\,}
\usepackage[dvips]{graphicx}
\usepackage{float}
\usepackage{epsf,graphics,graphicx}
\usepackage{rotating}
\usepackage{lscape}
\newcommand{\lsim}{\!\!\!\phantom{\le}\smash{\buildrel{}\over
 {\lower2.5dd\hbox{$\buildrel{\lower2dd\hbox{$\displaystyle<$}}\over
                                 \sim$}}}\,\,}
\newcommand{\gsim}{\!\!\!\phantom{\ge}\smash{\buildrel{}\over
{\lower2.5dd\hbox{$\buildrel{\lower2dd\hbox{$\displaystyle>$}}\over
                               \sim$}}}\,\,}

\def\h{$^{\rm h}$}
\def\m{$^{\rm m}$}

\begin{document}

 \title[ {\it e-}MERLIN observations at 5\,GHz of GOODS-N] { {\it e-}MERLIN observations at 5\,GHz of the GOODS-N region: 
pinpointing AGN cores in high redshift galaxies}
\author[D. Guidetti et al.]  {D. Guidetti \thanks{E-mail: d.guidetti@ira.inaf.it}$^1$, M. Bondi$^1$, I. Prandoni$^1$, R.J. Beswick$^2$, T.W.B. Muxlow$^2$,
\newauthor N. Wrigley$^2$, I.R. Smail$^3$, I. McHardy$^4$\\
   $^1$ INAF - Istituto di Radioastronomia, via Gobetti 101, I--40129 Bologna, Italy\\ 
   $^2$MERLIN/VLBI National Facility, Jodrell Bank Observatory,\\
   University of Manchester, Macclesfield, Cheshire, SK11 9DL, UK,\\
   $^3$Institute for Computational Cosmology, Durham University, Durham, DH1 3LE, UK,\\
   $^4$School of Physics and Astronomy, University of Southampton, Southampton SO17 1BJ, UK.
}

\date{Received }
\maketitle

\begin{abstract}

We present 5\,GHz {\it e-}MERLIN observations of the GOODS-N region at sub-arcsec resolution 
(0.2--0.5\,arcsec). These data form part of the early commissioning observations for the {\it 
e-}MERLIN interferometer and a pilot for the {\it e-}MERLIN legacy program eMERGE. A total of 17 
sources were detected with $S/N>3$. These observations provide unique information on the radio 
source morphology at sub-arcsec scales. For twelve of these sources, deeper 1.4\,GHz MERLIN+VLA 
observations at the same spatial resolution are available, allowing radio spectral indices to be 
derived for ten sources on sub-arcsec angular scales. Via analysis of the spectral indices and 
radio morphologies, these sources have been identified as AGN cores in moderate-to-high redshift 
(1$<z<$4) galaxies. These results have provided AGN (or AGN candidate) classification for six 
previously unclassified sources and confirmed the AGN nature of the rest of the sample. 
Ultimately the eMERGE project will image the GOODS-N 
region at 1.4 and 5\,GHz with higher resolution (about 50\,mas at 5\,GHz) and down to 
sub-$\mu$Jy sensitivities. The unique combination of sensitivity and spatial resolution will be 
exploited to study star formation and AGN activity in distant galaxies.

\end{abstract}
\begin{keywords}{galaxies: evolution -- galaxies: active  --
  galaxies: starburst -- cosmology: observations -- radio continuum:
  galaxies}
\end{keywords}

\section{INTRODUCTION}
\label{sect-intro}

The apparent relationship between the mass of a super-massive black hole (SMBH) and that of the 
pressure-supported spheroid hosting it (e.g. \citealt{Magorrian98}) is one of the most clear 
empirical indications of the potential importance of active galactic nuclei (AGN) and 
accretion-driven feedback mechanisms in galaxy formation and evolution theories. Indeed, the 
peak of QSO activity has been established at $z\sim 2$ (e.g. \citealt{Schmidt68}; 
\citealt{Hasinger05}), at the same epoch when star formation activity in the Universe was also 
at its peak. These processes often co-exist in galaxies, with observational evidence for the 
presence of an embedded AGN in ultraluminous starburst galaxies \citep{Alexander05,Alexander08} 
and in 20-30\% of $z\sim 2$ massive star-forming galaxies in the Great Observatories Origins 
Deep Surveys (GOODS, \citealt{Giavalisco04}) fields (\citealt{Daddi07,Mullaney12}).

It is an essential if galaxy formation theories are to be tested that an assessment of the 
importance of embedded AGN activity in star forming galaxies be made at moderate to high 
redshifts. As such we need to identify the contribution of obscured accretion activity in 
bolometric surveys of the evolution of star formation. In other words, we need to identify and 
separately trace the total (obscured and unobscured) star formation in the bulges of individual 
high-redshift galaxies, the related nuclear activity, and any star formation occurring on larger 
scales within a disk. Only by doing this will we be able to obtain a bolometrically complete 
census of star formation and of the growth of galaxies and their SMBHs. However, a complete 
census of both star formation and AGN galaxy activity, especially at high redshifts, is 
challenging due to dust extinction and gas obscuration by circumnuclear material.

Deep radio surveys provide a powerful, obscuration-independent tool for measuring {\it both} 
star formation and AGN activity in high-redshift galaxies, hence tracing the apparently 
simultaneous development of the stellar populations and the black-hole growth in the first 
massive galaxies. Indeed, multi-wavelength studies of deep radio fields show a composite 
population of star-forming galaxies and AGN (e.g. \citealt{Prandoni01}; \citealt{Ciliegi05}; 
\citealt{Afonso06}; \citealt{Smolcic08}; \citealt{Seymour08}; \citealt{Mignano08}; 
\citealt{Padovani09}), with the former dominating at the lowest flux densities 
($S<50-100$\,$\mu$Jy). Interestingly, about half of the AGNs probed in the deepest radio fields 
are characterised by relatively low radio-to-optical flux density ratios ($R\ll100$), as 
compared to classical radio-loud AGNs (see e.g. \citealt{Padovani09}). This is a direct evidence 
of radio-quiet AGNs being present at $\mu$Jy flux levels. Such a result supports the idea that 
radio-quiet AGNs are not necessarily radio silent (see \citealt{Kukula98}), and paradoxically 
opens the perspective of studying the entire AGN population, including the radio-quiet 
component, at radio wavelengths.

\subsection{Identifying AGN and Starburst Components}

Radio-quiet AGNs share many properties with star-forming galaxies: they have similar radio 
luminosities ($10^{22-24}$\,WHz$^{-1}$) and similar infrared/radio flux ratios. In addition 
radio-quiet AGNs are typically characterised by Seyfert-2-like optical spectra, which are often 
difficult to distinguish from those of star-forming galaxies (see e.g. \citealt{Prandoni09}). 
This makes it difficult to separate radio-quiet AGNs from star-forming galaxies, even with 
multi-wavelength information, although the availability of IRAC colors and/or X-ray data can 
help (\citealt{Prandoni09}; \citealt{DelMoro}).\\ One of the most direct ways to pinpoint 
embedded AGN cores in galaxies is the detection of a high surface brightness radio core through 
high resolution -- VLBI-like -- observations. In order to reliably separate radio structures on 
several scales, from AGN core/inner jets ($\ll$1\,kpc) to nuclear/disk starbursts ($\geq$1\,kpc) 
in high redshift ($z>1$) galaxies, sub-$\mu$Jy imaging sensitivity together with milli-arcsec 
resolution is clearly crucial (see e.g. \citealt{Kewley00}; \citealt{Garrett02}; 
\citealt{Biggs08}; \citealt{Chi13}; \citealt{Parra10}). Pioneering Multi-Element Radio Linked 
Interferometer Network (MERLIN) observations at sub-arcsec resolution of the Hubble Deep field 
North (HDF-N) field, included in the GOODS-N region, did show that below a flux density of about 
70\,$\mu$Jy, the majority of the radio sources consist of powerful star-forming galaxies, 
typically at $z<1.5$ \citep[][hereafter M05]{Muxlow05}. However a radio-AGN population was also 
morphologically identified, which had been classified as starbursts at all other wavelengths 
\citep{Casey09}. \\

\subsection{{\it e-}MERLIN Pilot Observations}

The {\it e-}MERLIN interferometer is a major upgrade of the existing MERLIN array working at 
centimetre wavelengths.  The array consists of up to seven radio radio telescopes spread across 
the UK and is equipped with a new optical fibre network and correlator, providing a ten-fold 
sensitivity boost, compared to MERLIN.

The {\it e-}MERLIN Galaxy Evolution Survey ({\it e}MERGE) Legacy project \citep{Muxlow08} 
represents the largest element of the recently approved {\it e-}MERLIN Legacy Programme, a set 
of large and high-impact surveys addressing fundamental questions in astronomy and astrophysics. 
The {\it e}MERGE Survey project aims to exploit the unique combination of
 {\it e-}MERLIN sensitivity (sub-$\mu$Jy rms noise) and spatial resolution (50\,mas at 6\,cm, 
comparable to that of the {\it Hubble Space Telescope}), to study star-formation and AGN 
activities in high redshift ($1<z<4$) galaxies, with special focus on possible co-existence and 
co-evolution of the two phenomena. One of two fields targeted by the project is the GOODS North 
(GOODS-N), the most intensively observed area on the northern sky, which includes the Hubble 
Deep Field North (HDF-N).

Here we present early {\it e-}MERLIN commissioning data at 5\,GHz of a 10\,arcmin field within 
the GOODS-N region at sub-arsec resolution in the framework of the {\it e-}MERGE project. One of 
the aims of these pilot observations is to pinpoint possible problems in the upgraded {\it 
e-}MERLIN system. The images are the first deep, sub-arcsecond {\it e-}MERLIN at 5\,GHz of such 
region. These 5\,GHz {\it e-}MERLIN observations can be considered as complementary to those 
previously obtained at 1.4\,GHz with MERLIN and presented by M05 who combined 
MERLIN and Very Large Array (VLA) images at 1.4\,GHz
for 92 radio sources with $S_{\rm 1.4} > 40$\,$\mu$Jy (extracted from the 
original 1.4\,GHz catalogue, see \citealt{Richards00}), and located within a region of $10\times 
10$\,arcmin$^2$ centred at R.A.\/~12\h~36\m~49\fs4, Dec.\/ +62\degr~12\arcmin~58\farcs00 
(J2000). These images (M05), were not corrected for the primary beam attenuation, have angular 
resolution in the range 0.2--0.5\,arcsec and reach a $1\sigma$ sensitivity of about 
3\,$\mu$Jy\,beam$^{-1}$. The M05 data have recently re-imaged over a large contiguous area and 
primary beam corrected (Wrigley et al., in prep) reaching a $1\sigma$ sensitivity of 
3.5\,$\mu$Jy\,beam$^{-1}$ at the pointing centre. 

The new pilot C-band study presented here is $\sim$5 times shallower than the previous L-band 
observations of M05. This, together with the lack of short spacings information, makes our pilot 
observations best suited to pinpoint relatively bright and compact sources associated to radio 
emitting AGNs, rather than probe faint and/or large-scale radio emission. Combining these pilot 
5\,GHz data with the new 1.4\,GHz images (Wrigley et al., in prep) we derived for the first time 
spectral indices at sub-arcsec resolution.

The paper is organized as follows. Sects.~\ref{sect-obs} and~\ref{sect-images} present the 
5\,GHz {\it e-}MERLIN commissioning observations, the data reduction and the source detection 
details. The spectral indices derived on sub-arcsec scale between the pre-existing 1.4\,GHz data 
and the new 5\,GHz ones are discussed in Sect.~\ref{sect-spectral}, together with the 
diagnostics used to identify AGN activity in the observed galaxies. Individual source details 
are given in Sect.~\ref{sect-details}, where the classification of each source is discussed. 
Conclusions and future perspectives are summarized in Sect.~\ref{future}.

\section{OBSERVATIONS AND DATA REDUCTION}
\label{sect-obs}		

As part of the commissioning activities of {\it e-}MERLIN, in July and August 2011 we observed a 
10\,arcmin diameter region within the GOODS-N field, centred at R.A.\/~12\h~36\m~40\fs0, Dec.\/ 
+62\degr~12\arcmin~47\farcs98 (J2000), for a total of 156 hours in the frequency range 
4.5-5\,GHz (hereafter referred to as 5\,GHz). This position differs by about 1\,arcmin from the 
pointing centre used by M05, but was chosen to maximise S/N at known source locations.

A sub-array of five of the seven {\it e-}MERLIN telescopes were used throughout these pilot observations, with the 
76-metre diameter Lovell and the 32m
Cambridge telescopes being unavailable at the time. 
The absence of the Cambridge antenna reduces the maximum spatial resolution of our 5\,GHz observations, making it comparable to that of the previous 1.4\,GHz MERLIN+VLA observations (see M05). Furthermore, the omission of the Lovell telescope gives rise to an approximately homogeneous array with 25m diameter elements, yielding a predictable primary beam.

The 5\,GHz data were taken in four adjacent $512\times0.25$\,MHz
intermediate frequency channels (IFs) and observed both circular polarizations, with correlator outputs integrated every second.
The observations were conducted with a cycle time of 1.5\,min
on the phase calibrator and 8\,minutes on the GOODS-N target 
yielding a total on-source integration time of 106\,hours.

\subsection{Calibration}
The data were inspected, flagged, calibrated and imaged 
using the NRAO {\sc aips} package, following standard procedures for
wide bandwidth observations.

Visibilities that were clearly corrupted by instrumental phase errors, 
telescope errors, or radio frequency interferences
were spectrally removed (channel by channel)
by using the tasks {\sc spflg} and {\sc ibled}.

All the calibrators and target data were then combined into a single multi-source
file 
and averaged down to $128\times1$\,MHz channels, considerably reducing the size of the dataset. 
The absolute flux-density scale was calibrated through 
observations of sources 3C\,286 (primary flux calibrator) and 1407+284 (OQ208,
secondary point-source calibrator) with flux density 
 boot-strapped to 3C\,286.
Assuming a point-source model and a spectrum within the observing band for OQ208
we established a flux density equal to 2.45\,Jy  
at the central observing frequency (\citealt{Baars} scale).\\
The flux calibrator 1407+284 was also used for bandpass calibration whereas phase correction of visibilities were calibrated by observing 
the calibrator source 
1241+603, located 2$^\circ$ away from the pointing centre.
The average
flux density for 1241+603 was measured to be 0.216$\pm$0.007\,Jy,
a value consistent with the flux reported by
the VLBA Calibrator List (NRAO)\footnote{
http://www.vlba.nrao.edu/astro/calib/} ($0.21$\,Jy at 5\,GHz).\\
The visibilities were then re-weighted
to take into account the different sensitivities of the telescopes
within the {\it e-}MERLIN array and finally time-averaged (4\,s), in order to further reduce the data volume and speed up the imaging process. 
The effects of frequency and time averaging on the derived source
fluxes are discussed in Sect.~\ref{sect-images}.

\subsection{Imaging}
We produced total intensity 
images of all 92 radio sources presented in M05. These were obtained by averaging the four available IFs (task {\sc imagr})
and restoring clean components with a circular beam of 0.5\,arcsec diameter. Image dimensions measure 128$\times$128
pixels and a cellsize equal to 0.0625\,arcsec was chosen generating fields some 8\,arcsec across.
Whenever a detection was found within one of the 0.5\,arcsec resolution fields, 
a higher resolution (0.2\,arcsec) map was re-imaged. The only exception is source J123725+621128, 
a relatively extended double radio source (see last panel of Fig.~\ref{fig:images} and source details in Sect.~\ref{sect-details}), 
where higher resolution imaging was not attempted in order to probe the extended emission visible at 1.4\,GHz (M05).
We also imaged 10 unresolved sources 
from the original 1.4\,GHz VLA catalog produced by \citet{Richards00}
(see also \citealt{Biggs06,Morrison10})
with arcsec-scale flux density $> 300$\,$\mu$Jy, 
located within a distance of 10\,arcmin from our field centre at 0.2\,arcsec resolution. 
These sources lie in a region outside the usable primary beam of MERLIN so were not included in M05.

All the images were cleaned until the peak of the residuals was about three 
times the rms noise ($\sigma$). Cleaning deeper risks so-called ``clean bias'', that may affect deep
imaging in presence of poor UV coverage (\citealp{White97}; \citealp{Condon98}). 
Frequency channels at the edges of the passband are generally characterized by very low 
sensitivity, therefore the lowest rms noise level in the final images is obtained by excluding these
 channels. This results in an effective bandwidth of 120\,MHz within each IF.

The average off-source noise level, measured before correction for the primary beam, in our 
naturally-weighted images at 0.2\,arcsec and 0.5\,arcsec resolution is $\simeq15$ and 
$\simeq20\,\mu$Jybeam$^{-1}$ respectively. From image analysis we verified that the noise is close 
Gaussian, however, the values reported above are about 2 times higher than expected from thermal noise. The 
cause of sensitivity degradation in these early {\it e-}MERLIN commissioning 5\,GHz observations have been 
identified with elements of the signal path introducing noise in the data signal. Following these 
observations, additional first stage amplification has been introduced to mitigate this effect.

The images were corrected for the primary beam attenuation, adopting the parameters used for the VLA antenna at 5\,GHz. This is a reasonable assumption since all the antennas
participating to our {\it e-}MERLIN observations are effectively 25-metre dishes.
The primary beam correction results in rms noise which increases as a function of distance from the pointing phase centre, varying from 
approximately  20-30\,$\mu$Jy within 200\,arcsec of the pointing centre, to 40-50\,$\mu$Jy at 300\,arcsec before rapidly growing to several hundred micro-Janskys.

\begin{table*}
\caption{Radio sources detected by {\it e-}MERLIN at 5\,GHz.
 Col.\,1: source name; Col.\,2: distance from the pointing centre; Col.\,3: 
local rms noise after the primary beam correction;
 Col.\,4: Right Ascension and Declination of the peak;
Col.\,5: signal-to-noise ratio; Col.\,6 \& 7: integrated flux densities at 5\,GHz and 1.4\,GHz, respectively;
 Col.\,8: redshift. All parameters refer to 0.2\,arcsec resolution images, with the exception of source J123725+621128, for which the resolution is 0.5\,arcsec. 
In italic are the sources with uncertain flux density due to the large primary beam and smearing corrections.
\label{tab:images}}
\begin{tabular}{lccccrccc}
\hline
source & d & $\sigma_{\rm local}$ & \multicolumn{2}{c} {position}  & $S/N$ &  S$_{5}$ & S$_{1.4}^{*}$ & $z$ \\
       & [arcsec]  & [$\mu$Jy] & RA & DEC &  & [$\mu$Jy\,beam$^{-1}$]  & [$\mu$Jy\,beam$^{-1}$]   & \\
\hline
J123606+620951 & 293 & 40 & 12 36 06.60 & 62 09 51.16 & 4.0 & 161 & 302 & 0.6379$^1$ \\
J123617+621540 & 233 & 28 & 12 36 17.55 & 62 15 40.77 & 5.1 &  140 & 176 &1.993$^2$  \\
J123623+621642 & 261 & 35  & 12 36 23.54 & 62 16 42.80 & 3.7 & 130 & 437 &1.918$^2$  \\
J123624+621743 & 314 & 45  & 12 36 24.82 & 62 17 43.67 & 3.3 & 150 & 230 &-      \\
{\it J123640+622037} & 470 & 234 & 12 36 40.14 & 62 20 37.69 & ~4.7 & $\sim$1100 &  &0.8$^3$  \\
J123642+621331 & 46  & 16  & 12 36 42.09 & 62 13 31.44 & 5.7 & 93 & 259 &4.424$^4$  \\
J123644+621133 & 81  & 17 & 12 36 44.39 & 62 11 33.16 & 21.6 & 543 & 722 & 1.050$^5$  \\
J123646+621404 & 89  & 18  & 12 36 46.33 & 62 14 04.70 & 5.7 & 201 & 168 &0.961$^1$  \\
{\it J123649+620438} & 492 & 303 & 12 36 48.97 & 62 04 38.69 & ~3.3 & $\sim$1000 &  &0.113$^6$ \\
J123649+620737 & 317 & 51 & 12 36 49.65 & 62 07 38.10 & ~3.3 & 169  &  & 2.315$^{7,8}$ \\
J123650+620844 & 254 & 31 & 12 36 50.20 & 62 08 44.76 & 3.5 & 109  & 160 & 0.434$^1$  \\
J123652+621444 & 147 & 20  & 12 36 52.89 & 62 14 44.08 & 4.1 & 83 & 66 & 0.321$^1$  \\
J123659+621832 & 369 & 67 & 12 36 59.33 & 62 18 32.69 & 12.6 & 850  & & - \\
J123714+620823 & 361 & 71  & 12 37 14.94 & 62 08 23.21 & 18.3 & 1305  & 1369 & -      \\
{\it J123721+620708} & 447 & 194 & 12 37 21.40 & 62 07 08.24 & ~3.1 & $\sim$600  & &       -  \\
J123721+621129 & 299 & 47 & 12 37 21.25 & 62 11 29.96 & 6.7 & 300 & 199 & 1.56$^9$ \\
J123725+621128 & 331 & 49 & 12 37 26.04 & 62 11 28.67 & 3.8 & 187 & 546 & -      \\
\hline
\multicolumn{9}{l}{$^{*}$ flux densities from primary beam corrected MERLIN+VLA data (Wrigley et al. in prep.)}\\
\multicolumn{9}{l}{$^1$ \citet{Cowie04}; $^2$ \citet{Chapman04}; $^3$ \citet{Donley05}; $^4$ \citet{Waddi99}; $^5$ \citet{Cohen00};}\\
\multicolumn{9}{l}{ $^6$ \citet{Horn05}; $^7$ \citet{Smail04}; $^8$ \citet{Casey09}; $^9$ \citet{Barger02}}\\
\end{tabular}
\end{table*}

\begin{figure*}
\centering
\includegraphics[width=18.5cm]{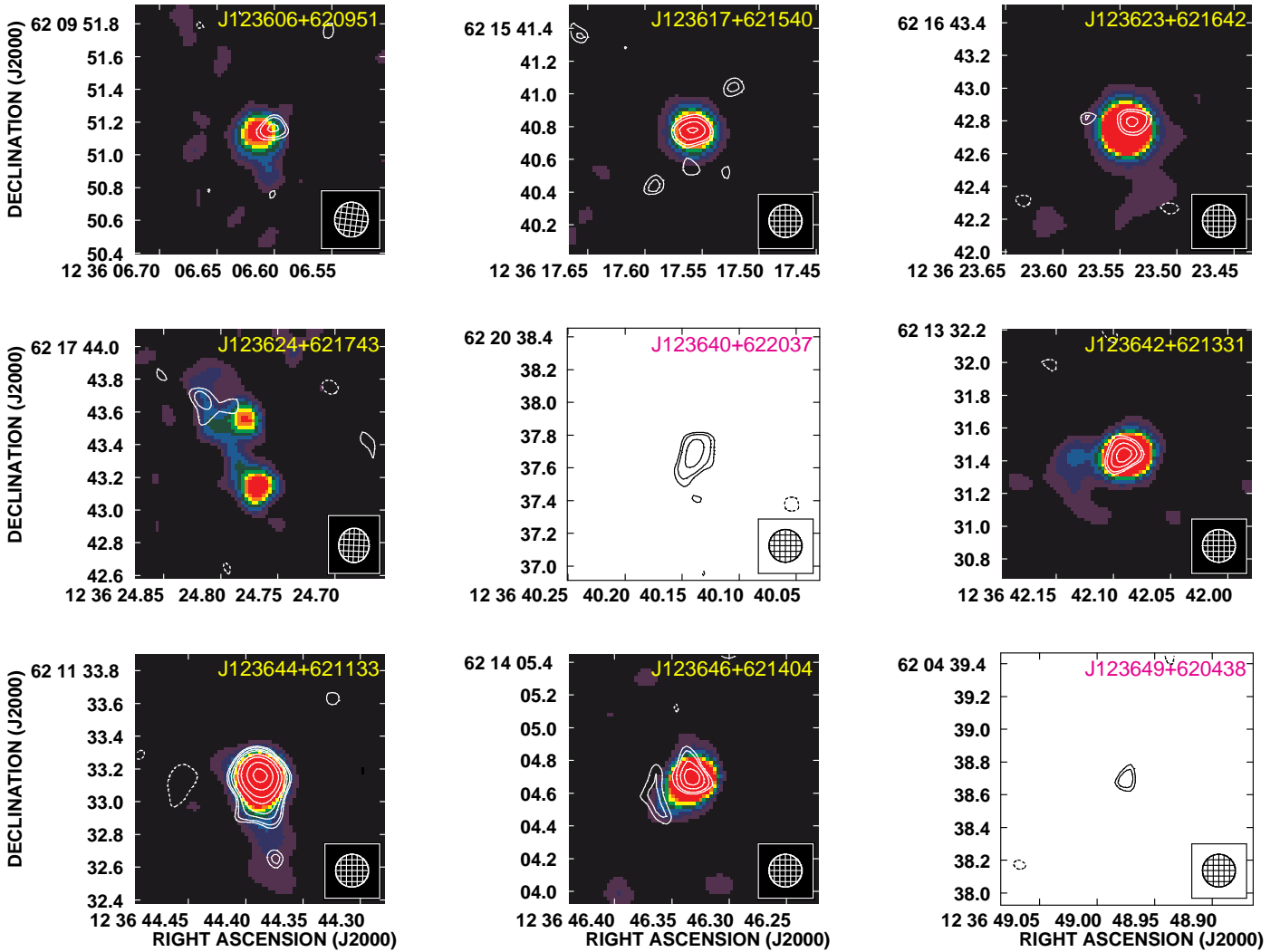}
\caption[]{Radio contours of the sources detected at 5\,GHz overlaid to 
the 1.4\,GHz emission obtained by the combined MERLIN-VLA data (Wrigley et al., in prep).
The images with blank backgrounds do not have 1.4\,GHz MERLIN-VLA data available.
The restoring beam is 0.2\,arcsec for all sources.
The contours are drawn at -2.5, 2.5, 3, 4, 5, 7, 10, 15, 20 times the local rms noise listed in
Table~\ref{tab:images}.
\label{fig:images}
}
\end{figure*}

\begin{figure*}
\addtocounter{figure}{-1}
\centering
\includegraphics[width=18.5cm]{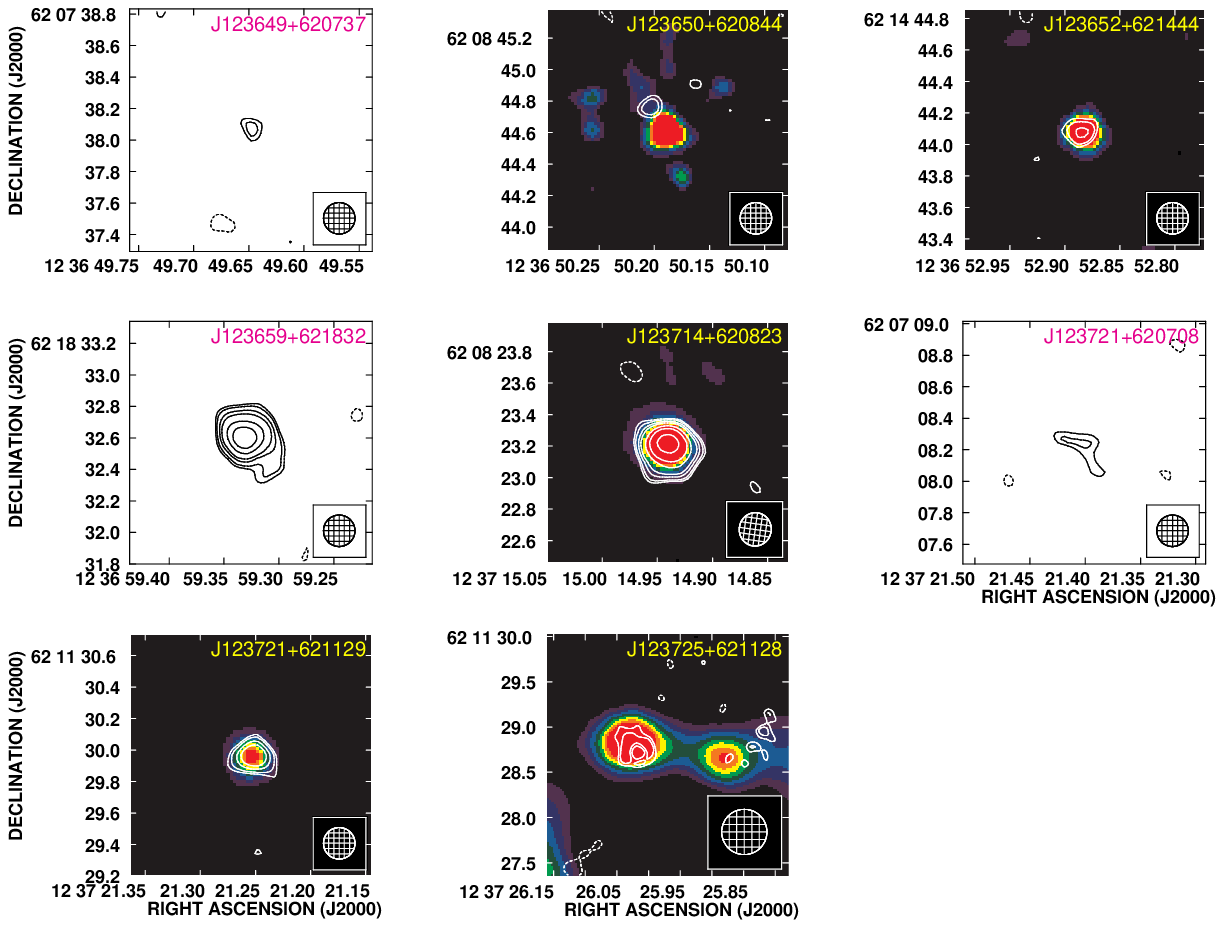}
\caption[]{(continued). Radio contours of the sources detected at 5\,GHz overlaid to 
the 1.4\,GHz emission obtained by the combined MERLIN-VLA data (Wrigley et al., in prep).
The images with blank background do not have 1.4\,GHz MERLIN-VLA data available.
The restoring beam is 0.2\,arcsec for all sources, except for source J123725+621128, where it is 0.5\,arcsec. 
The contours are drawn at -2.5, 2.5, 3, 4, 5, 7, 10, 15, 20 times the local rms noise listed in
Table~\ref{tab:images}.
}
\end{figure*}

\section{SOURCE DETECTION}
\label{sect-images}

At a 3$\sigma$ threshold,
a total of 17 sources were detected (12 from the M05 sample and 5 from the \citet{Richards00} catalogue).
As expected, all these sources have relatively high measured flux densities ($> 80-100$\,$\mu$Jy), 
and all but two are consistent with being unresolved.
In contrast, all the M05 sources were resolved at 1.4\,GHz.
This is easily explained by the lack of short
baselines in our 5\,GHz {\it e-}MERLIN observations, together with a factor of
$\sim 5$ worse sensitivity than for the 1.4\,GHz observations (coupled with the fact that extended synchrotron emission is typically characterized by a declining spectrum with frequency).

The 5 sources from the \citet{Richards00} catalogue
are located at larger angular distances from the phase centre and are brighter 
than those in the M05 sample.
 
The measured source parameters are presented in Table~\ref{tab:images} which include the distance from the pointing centre, the local rms noise value, the peak position, the signal to noise  ratio (S/N), the integrated flux density at 1.4 and 5\,GHz.
The source redshift is included, where available.
For unresolved sources, the integrated flux density is equal to the peak
brightness, obtained through a quadratic polynomial fitting ({\sc aips} verb {\sc maxfit}),
whereas for the two extended sources, the  integrated  spectral flux density is derived by integrating 
over a circular region with radius of 5 pixels centred on the peak flux position.
The peak brightness of the sources is corrected for the combined bandwidth and time smearing effects following \citet{BS89} which predicts that sources within a distance of 300\,arcsec from the pointing centre (imaged with a resolution of 0.2\,arcsec) have intrinsic peak brightnesses reduced by 
up to $10\%$, while at $\sim400$\,arcsec the smearing reduces the peak flux density by up to $\sim 30\%$.
These smearing corrections, in combination with primary beam uncertainties, yields highly uncertain flux density measurements for sources lying beyond 400\,arcsec  
from the pointing centre. In these cases we only provide a rough estimate of their flux densities which, 
being all unresolved, are set equal to their peak flux density measurements.

Contour images at 5\,GHz of the 17 detected sources are shown in Fig.~\ref{fig:images}.
The radio contours are overlaid on the pre-existing 1.4\,GHz images 
obtained at the same resolution.
As mentioned earlier, the 5 detected sources selected from the VLA catalog by \citet{Richards00} were not imaged by M05. 
Therefore, for such sources we do not have a sub-arcsec 1.4\,GHz image.
 
In general the source peak positions 
at 1.4 and 5\,GHz are coincident within the errors,
suggesting a common origin of the two emissions. 
The two most notable exceptions are sources J123624+621743 and J123650+620844 
where a significant offset is found between the 1.4 and 5\,GHz emission centroids. 
These particular sources will be discussed further in Sect.~\ref{sect-spectral}.

Figure~\ref{fig:noise} shows our source detection expectation based on 
the flux densities of the sources at 1.4\,GHz used here as a proxy of those expected 5\,GHz, 
under the assumption that the dominant component of the radio emission is a compact, flat spectrum ($\alpha=0$) AGN core. 
Such fluxes have to be compared to our $3\sigma$ detection threshold.
It is clear that only the brightest sources in the M05 sample can be in principle detected. 
This is mainly a consequence of the fact that in our maps the achieved noise is, as already mentioned in
Sect.~\ref{sect-obs}, a factor two higher than expected.

\section{PINPOINTING AGN CORES IN GOODS-N GALAXIES}

As discussed in Sect.~\ref{sect-intro}, a clear separation between AGN and star-formation processes in galaxy samples can be very difficult, especially at high redshifts, where gas/dust obscuration can be very important and composite systems (where both phenomena co-exist) are likely to be more frequent. One of the safest way to classify extragalactic radio sources as AGN or starburst is through optical spectroscopy. Unfortunately, these kind of observations 
can be extremely time consuming or not possible at all due to the faintness of 
the targets. Moreover even spectral classification is not always clear, especially in presence of low luminosity Seyfert-2 AGNs, whose spectra are not always easily distinguished from those of star-burst galaxies. 

\subsection{AGN/Starburst discrimination methodology}
A multi-wavelength approach is typically used when attempting to separate AGN from star-forming galaxies, possibly taking advantage of all available diagnostic tools.
Each diagnostic test, taken by itself, is usually not sufficient to securely
classify galaxies, but when two or more criteria can be verified, the classification can be considered more reliable.

The following diagnostics
have been used by M05 to classify the 92 sources in their sample:

\begin{enumerate}

\item
The radio-FIR correlation -- this correlation is remarkably tight for star-forming galaxies and holds over a wide range of luminosities \citep[e.g.][]{deJong85, Condon92, Beswick08}
and redshifts \citep[e.g.][]{Garrett02}. However the radio-FIR correlation is not useful to distinguish star-forming galaxies from radio-quiet AGNs (see discussion in Sect.~\ref{sect-intro}).

\item
The X-ray luminosity -- low redshift starburst galaxies are soft-band ($< 2$ keV) dominated sources (\citealt{Ptak99}, \citealt{Alexander02}) with X-ray luminosities $L_X< 10^{35}$~W.  
X-ray luminosities $L_X> 10^{35}$~W,  on the other hand, usually indicate the presence of an AGN \citep{Richards07}.
However, exceptions exist, like e.g. low luminosity radio galaxies with X-ray luminosities
in the range $L_X \sim 10^{33} - 10^{35}$~W \citep{Evans06}.
\item
The radio morphology -- jets/lobes emission is a clear indication of the 
presence of a radio AGN. A sub-arcsec-scale bright compact core can be 
associated to an AGN, although the possibility of a compact starburst can not be excluded at a resolution 
of 0.2\,arcsec (corresponding to $\sim 1-1.5$\,kpc at $1<z<4$, see  \citealt{Muxlow05,Richards07}; \citealt{Batejat12} for Arp 220; \citealt{Bondi12} for Arp 299). The optical appearance, if observations with enough 
resolution are available, can be usefully coupled to the radio morphology 
to discriminate between an AGN or starburst origin of the radio emission.

\item
The radio spectral index -- radio emission with spectral index $\alpha\lsim 0.4$ ($S\sim \nu^{-\alpha}$) is likely to be 
powered by an AGN. On the other hand, steeper-spectrum emission can be produced 
by either radio jets/lobes (and therefore still AGN dominated) or by supernova remnants (starburst 
dominated). Inspecting the radio morphology with an adequate resolution can
allow discrimination between these two possibilities.
\end{enumerate} 

By applying these diagnostic tools, a classification as AGN/AGN-candidate or SB/SB-candidate was proposed 
by M05 for most ($\sim 82\%$) of their 92 radio sources. Nine additional sources were later classified by 
\citet{Richards07}.

As discussed in Sect.~\ref{sect-intro}, one of the most direct ways to pinpoint embedded AGN cores in galaxies is the detection of a flat-spectrum high surface brightness self-absorbed radio core  through high resolution multi-frequency observations. In this paper, we take advantage of our pilot sub-arcsec 5\,GHz {\it e-}MERLIN observations, to better assess the presence of AGN cores in GOODS-N galaxies. By combining our 5\,GHz maps with the recently re-{\sc clean}ed and primary beam corrected MERLIN+VLA data at 1.4\,GHz (Wrigley et al. in prep),
obtained at same resolution, we derived 1.4--5\,GHz spectral indices for 10 of the 17 detected sources.
Spectral index information and radio morphologies were both used to confirm previous AGN classification, when available, and to provide a new one for the sources still unclassified. 
The spectral index information is particularly relevant here, first because the radio morphology alone does not necessarily  allow us to discriminate a compact starburst from an AGN core (see discussion above, item iii). Secondly, brightness temperature measurements 
would not help us in the source classification, as the spatial resolution of our observations 
is not high enough to be sensitive to excess of 10$^5$\,K. 
Indeed, the angular resolutions of our maps (0.2-0.5\,arcsec FWHM) give brightness temperatures in a range (700-5000\,K).

Our classification also takes advantage of any pre-existing multi-wavelength information, in line with those used by M05 and \citet{Richards07}. 

Spectral index analysis of M05 (see item iv above) was based on 1.4 and 8.5\,GHz radio fluxes obtained from arcsec resolution images (3.5\,arcsec, \citealt{Richards00}), whereas our spectral indices are instead derived between 1.4 and 
5\,GHz sub-arcsec resolution flux densities. The latter are therefore better suited to pinpoint flat-spectrum compact AGN cores.

For the unresolved sources at 5\,GHz, the spectral 
index is derived using the peak flux density of the corresponding 
1.4\,GHz component, when the peaks at the two frequencies overlap. 
For the two sources (J123624+621743 and J123650+620844)
with a significant positional offset between the peaks at the two
frequencies (see discussion in Sect.~\ref{sect-images}) we do not provide a spectral index.
For the two extended sources at 5\,GHz (J123644+621133 and
J123646+621404) we derived the total flux at the lower frequency 
integrating over the same region used to derive the flux density at 5\,GHz.
It is worth noting that for source J123725+621128 we detect only the emission
from a region eastwards the core, and therefore the spectral index refers to this area.

One caveat must be stated clearly: whereas the 1.4 and 8.5\,GHz observations
were made relatively close in time (1-2\,years), our {\it e-}MERLIN
5\,GHz observations are separated from the 1.4\,GHz 
data by more than 10 years. Variability
in the radio domain may therefore be an issue and should be considered in the interpretation of our spectral indices.

\section{RESULTS}
\label{sect-spectral}

The derived spectral indices are presented in Table~\ref{tab:spix}, together with the previous determination between 1.4 and 8.5\,GHz at arcsec resolution \citep{Richards00}.
For sources not previously 
imaged at sub-arcsec scale by M05, we derive a spectral index using the arcsec-scale 1.4\,GHz (and 8.5\,GHz, where available)  flux density measurements 
(\citealt{Richards00}) and our sub-arcsec 5\,GHz flux densities. Such spectral indices have to be considered as tentative, since sub-arcsec 5\,GHz 
flux densities may be underestimated with respect to those obtained at lower resolution. For reasons of conservatism, such spectral index values are not included in Table~\ref{tab:spix}.

In general there is a reasonable agreement between the two spectral indices, although we notice a tendency for our 1.4-5\,GHz spectral indices to be flatter than 
those derived between 1.4 and 8.5\,GHz. This is expected, as sub-arcsec resolution spectral indices are likely to better probe self-absorbed AGN cores 
than those derived at arcsec resolution. This supports multi-frequency sub-arcsec radio imaging as a promising technique to pinpoint AGN cores in high-redshift galaxies.

\begin{table}
\small\addtolength{\tabcolsep}{-4pt}
\caption{Spectral index and source classification comparison.
Col.\,1: source name.
Col.\,2: previous radio classification based on sub-arcsec radio imaging.
Col.\,3: spectral index between 1.4 and 8.5\,GHz \citep{Richards00}.
Col.\,4: spectral index between 1.4 and 5\,GHz.
Col.\,5: new classification.
The 5 sources listed at the bottom do not belong to the M05 sample 
and do not possess a classification based on sub-arcsec scale imaging.
\label{tab:spix}}
\begin{tabular}{ccrrc}
\hline
Source & class$^a$ & $\alpha^{1.4}_{8.5}$ & $\alpha^{1.4}_{5.0}$ & class$^b$\\
       &           &                     &            &         \\
\hline
J123606+620951 & AGN &  $\ge$ 0.56 $\,\,\,$ & 0.53$\pm$0.23   & AGN    \\
J123617+621540 & uncl &  $\ge$ 0.55 $\,\,\,$ & 0.19$\pm$0.22    & AGN?  \\
J123623+621642 & AGN & 0.63$\pm$0.07 & 1.02$\pm$0.23  & AGN   \\
J123624+621743 & uncl &               &          &   -     \\
J123642+621331 & AGN+SB? & 0.94$\pm$0.06 & 0.86$\pm$0.18   & AGN+SB?\\
J123644+621133 & AGN & 0.30$\pm$0.05   & 0.24$\pm$0.11   & AGN \\
J123646+621404 & AGN & -0.04$\pm$0.06 & $-0.15\pm$0.20  & AGN  \\
J123650+620844 & SB   &     $\ge$ 0.80 $\,\,\,$         &               &  -   \\
J123652+621444 & AGN & $0.12\pm$0.07 & $-0.19\pm$0.25  & AGN   \\
J123714+620823 & AGN & $0.15\pm$0.08 & 0.04$\pm$0.09   & AGN  \\
J123721+621129 & AGN & $-0.28\pm$0.06 & $-0.35\pm$0.22  & AGN  \\
J123725+621128 & AGN & 1.35$\pm$0.06 & 0.90$\pm$0.22    & AGN  \\
       &           &                     &         &            \\
J123640+622037 & AGN  &               &               & AGN    \\
J123649+620438 & - &              &               & AGN     \\
J123649+620737 & - & 0.56$\pm$0.07 &   & AGN?                \\
J123659+621832 & -  & 0.26$\pm$0.07 &       & AGN            \\
J123721+620708 & -   &               &               & AGN?    \\
\hline
\multicolumn{5}{l}{$^a$ \citet{Garrett01}, M05, \citet{Donley05},}\\
\multicolumn{5}{l}{ \citet{Richards07}, \citet{Chi13}; $^b$ This work. } \\
\end{tabular}
\end{table}

\begin{figure}
\centering
\includegraphics[width=8cm]{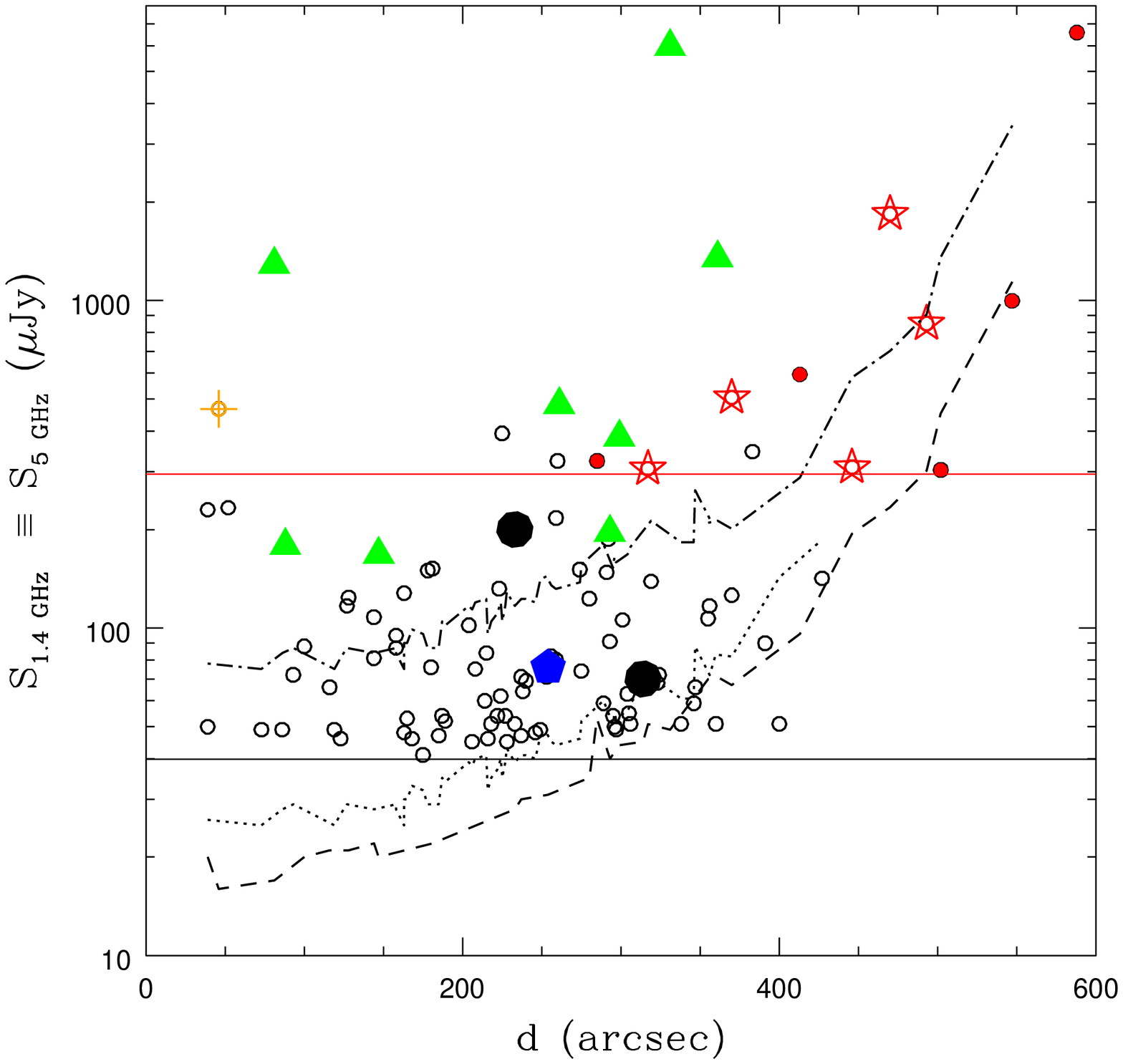}
\caption[]{
1.4\,GHz flux density versus distance from the pointing centre for the 92
sources imaged by M05 and \citet{Richards00}.
Undetected sources are shown as black empty circles for the M05 sample and red filled circles 
for \citet{Richards00} sources.
Detected sources are indicated by symbols in
different colors corresponding to different source classifications (from M05 or \citet{Richards00}, 
see discussion in Sect.~\ref{sect-spectral}): AGN (green triangles); starburst (SB, blue pentagons); possible composite 
source (AGN+SB?, orange pointed crosses); 
unclassified source (black circles and red stars for M05 and \citet{Richards00} sources, respectively). 
The black horizontal line indicates the flux density limit of the M05 sample (92 sources with $S_{(\rm{1.4\,GHz})}>40$\,$\mu$Jy), whereas the red horizontal line indicates the flux density limit of the additional 10 sources from the 
\citet{Richards00} catalogue imaged in the framework of this work ($S_{(\rm{1.4\,GHz})}>300$\,$\mu$Jy). 
The dashed and dotted lines represent the rms measured in the  5\,GHz images 
we produced at 0.2 and 0.5\,arcsec resolution, respectively. 
The dot-dashed line represents our $3\sigma$ detection threshold. 
\label{fig:noise}
}
\end{figure}

Table~\ref{tab:spix} summarises the pre-existing source classifications based on sub-arcsec radio imaging obtained by M05 and \citet{Richards07} 
and, where available, on higher resolution VLBI imaging (\citealt{Garrett01}; \citealt{Chi13}).  The additional five
 sources not belonging to the M05 sample (listed at the bottom) do not therefore have such classifications 
available. As expected, eight of our sources are associated with galaxies classified as AGN or AGN 
candidates by M05 or \citet{Richards07} (Fig.~\ref{fig:noise}). One of the additional sources was 
classified as a possible composite starburst+AGN object (Fig.~\ref{fig:noise}). Six sources were also 
detected by higher resolution Very Long Baseline Interferometry (VLBI) observations at L-band 
\citep{Garrett01,Chi13}, confirming the presence of an AGN core at the very centre of the galaxy. Only one 
source was previously classified as a starburst galaxy. Our new classification is discussed for each source 
in Sect.~\ref{sect-details} and the results are summarized in the last column of Table~\ref{tab:spix}.

We detect sources preferentially classified as AGN/AGN- candidates rather than starburst components which, 
as mentioned earlier (see Sect.~\ref{sect-images}), is a consequence of the fact that our 5 GHz 
observations are sensitive only to the bright compact radio sources in the observed field, and of the fact 
that AGNs are detected in large fractions at the brighter end of sub-mJy radio samples (78$\%$ at $S_{\rm 
1.4 GHz}>500\,\mu$Jy, \citet{Mignano08}. Indeed of the 19 sources in our sample $S_{\rm 1.4 
GHz}>300\,\mu$Jy, 10 of the 11 detected are positively identified as AGN/AGN candidates 
(Fig.~\ref{fig:noise}). Under the conservative assumption that the undetected sources are all lower surface 
brightness star-forming galaxies, we can conclude that at such flux densities $\sim 53\%$ (10/19) of the 
sources are AGN-dominated. Going to lower flux densities ($100<S_{\rm 1.4 GHz}<300\,\mu$Jy), and limiting 
our statistical analysis to sources within $d\leq 300$ arcsec from the image centre, where our detection 
threshold is sensitive enough (see Fig.~\ref{fig:noise}), we detect only 4 additional sources (out of 19 in 
these flux and distance intervals). Three of such sources are associated to AGN/AGN-candidates, 
corresponding to $\sim 16\%$ of the total (3/19). This has to be considered as a conservative estimate of 
the AGN fraction, as we cannot exclude, in this flux interval, to have steep-spectrum AGN among the 
undetected sources. Nevertheless our results are consistent with a decreasing fraction of AGNs with 
decreasing flux density, as found in wider arcsec-scale sub-mJy samples (see e.g. \citealt{Seymour08}).

\subsection{Source classification}
\label{sect-details}

In this subsection we present the updated classifications
based on our 5\,GHz sub-arcsec imaging and spectral index analysis of each source. Each
classification is discussed with respect to previous
indications from M05 and other multiwavelength observations.


{\bf J123606+620951}:
This source was unclassified by M05 and later classified as AGN by
\citet{Richards07}, thanks to new X-ray band information.
The integrated flux density measured directly from the $0.2\arcsec$ resolution 1.4\,GHz images is higher than that 
reported in Table A2 by M05 and derived from VLA only observations.
It is possible that the source is variable.  
The core position at 5\,GHz agrees within the errors with that were reported by 
M05 at 1.4\,GHz. The spectral index
we obtain is consistent with the arcsec-scale previous determination between 1.4 and 8.5\,GHz, and is not particularly informative ($\alpha\simeq 0.5\pm 0.2$). However the detection of a compact, possibly variable, component likely confirms the AGN origin of the radio emission.

{\bf J123617+621540}:
M05 observed a compact structure together with
two-sided emission extended over about 1.3\,arcsec, but
the source was unclassified because the radio spectrum was poorly constrained ($\alpha\ge 0.55$).
However, we clearly detect
a bright core with flat spectral index
($\alpha\sim 0.2\pm 0.2$), strongly suggesting the presence of an AGN.

{\bf J123623+621642}:
The source was classified in M05 as an AGN candidate. The presence of an AGN core was later confirmed  by global VLBI observations at 1.4\,GHz, that 
revealed a compact emission \citep{Chi13}.
At 5\,GHz we detect a compact component coincident with the peak
of the 1.4\,GHz radio source. 
The spectral index we derive is 
steeper ($\alpha\simeq 1.0\pm 0.2$) than that found between
1.4 and 8.5\,GHz, suggesting possible flux density variability
between the 1996-97 1.4 and 8.5\,GHz observations and our 2011
5\,GHz observations. 
The compact structure and the possible variability
both confirm the AGN origin of the radio emission. 
 
{\bf J123624+621743}:
The source does not have a  $1.4-8.5$\,GHz spectral index measurement. It 
has a faint X-ray counterpart but the lack of redshift information does
not allow a derivation of the X-ray luminosity.
We detect a weak, compact structure at 
0.2\,arcsec resolution with S/N of 3.3, that is not coincident with the peak
of the 1.4\,GHz image, but overlaps to its northern 
extended radio emission. 
To check the reliability of this detection, we used the Gaussian statistics
to evaluate the probability to have a noise spike in the 5\,GHz image above 3.3$\sigma$ within the area covered by the 
1.4\,GHz VLA emission. This assumption is motivated by the fact that the rms noise distribution is very close to Gaussian.
The probability for a spurious signal is of about 0.7\%, however
deeper observations are needed to confirm this detection.

Because of this offset, we have not calculated the spectral index. 
The source is associated with a relatively faint 1.4\,GHz source ($S_{(\rm{1.4\,GHz})}\sim70\,\mu$Jy)
and was not expected to be detected at 5\,GHz 
(it is well below our 5\,GHz detection threshold in Fig.~\ref{fig:noise}).
The presence of a significant positional offset between the two peaks at
1.4 and 5\,GHz and the apparent flux inconsistency make this detection uncertain.
Deeper observations are therefore needed to confirm this detection.

{\bf J123640+622037}:
This is one of the sources not previously imaged at 1.4\,GHz by M05, and therefore a classification was not previously provided.
However this object shows a radio-excess with respect to the radio/infrared correlation for radio-quiet AGNs and star-forming galaxies and was therefore classified as a radio-loud AGN \citep{Donley05}. This source is neither detected by {\it Chandra} nor has a $1.4-8.5$\,GHz spectral index determination. 
At 5\,GHz we observe a bright compact core  with a highly uncertain flux density
due to the large primary beam correction.
With the caveat of this uncertainty, we tentatively obtain a radio spectral index between 1.4 and 5\,GHz of $\sim 0.4$. The spectral index may be flatter, if we consider that some of the 5\,GHz flux could be resolved out in our sub-arcsec 5\,GHz images. This seems to confirm the AGN classification inferred by the high radio/infrared flux ratio.

{\bf J123642+621331}:
\citet{Waddi99} argued that this source is most likely
a distant ($z=4.424$) dusty starburst with a weak embedded AGN.
At 1.4\,GHz the source shows both diffuse and compact structures (M05).
The latter has been revealed at higher resolution by VLBI observations at 1.4\,GHz 
\citep{Garrett01,Chi13}, confirming the presence of AGN emission.
At 5\,GHz, we detect a compact component coincident with the peak of the 1.4\,GHz emission.
Since at 1.4\,GHz the radio source is extended, 
we use only the flux of the compact emission at 1.4\,GHz to derive a 
spectral index of $\alpha\simeq 0.9\pm 0.2$. This value is consistent with the spectral index derived at arcsec-scale between 1.4 and 8.5\,GHz. The steepness of the spectrum may confirm the composite nature of the source, and/or be explained by steepening effects introduced by the (large) $k$-correction of this high redshift source.

{\bf J123644+621133}:
This is a FR\,I radio galaxy. 1.4\,GHz sub-arcsec scale imaging shows a bright compact core and a short jet extending southwards  (M05).
1.4\,GHz VLBI observations \citep{Garrett01,Chi13} detected the compact core component.
A core+jet morphology is observed also in our 5\,GHz image, confirming the AGN nature of the source. Considering the same area emitting at 5 and 1.4\,GHz, we derive a rather flat spectrum with $\alpha\simeq 0.2\pm 0.1$. 

{\bf J123646+621404}:
An unresolved core has been detected by VLBI observations at 1.4\,GHz 
\citep{Garrett01,Chi13}. Extended emission southeast
of the nucleus was detected at 1.4\,GHz by M05. 
At 5\,GHz we observe a compact core and again some extended emission southeast
of the nucleus. The radio spectrum is inverted on both arcsec and sub-arcsec scales, confirming the AGN nature of the radio emission.
 
{\bf J123649+620438}:
This is one of the sources not previously imaged at 1.4\,GHz by M05 which does not have a $1.4-8.5$\,GHz spectral index determination. The AGN/SB discriminators in this source are ambiguous. The galaxy is detected by {\it Chandra} with a 0.5-8
keV X-ray luminosity of $L_X\simeq 5.3\times 10^{33}$\,W, and 
a rather soft-band dominated emission, $\Gamma = 1.6$, that would be consistent 
with starburst emission \citep{Alexander02}.
On the other hand, on the basis of an optical spectrum showing the H$\alpha$ line, 
the galaxy was classified as an AGN \citep{Horn05}.
We detect a bright core at 5\,GHz, that implies a flat or slightly inverted radio spectrum between 1.4 and 5\,GHz ($\alpha\sim -0.1$), supporting the low luminosity AGN scenario.

{\bf J123649+620737}:
This is one of the sources not previously imaged at 1.4\,GHz by M05. The overall radio spectral index on arcsec-scale is steep ($\alpha\sim 0.56\pm0.07$, see Table~\ref{tab:spix}), and the 1.4\,GHz arcsec-scale image shows extended emission.
The radio source is associated with a hot-dust dominated ultraluminous galaxy thought to host both an AGN and intense star forming regions \citep{Casey09}. Alternatively, the extended 1.4\,GHz emission may be AGN related extended emission and the 70\,$\mu$m emission may be due to a dusty torus surrounding the AGN. 
The compact component detected at 5\,GHz overlaps with the compact bright optical nucleus visible in the HST image (see Figure 3 in \citealt{Casey09}). Using our 5\,GHz flux density measurement we find a possible steepening of the radio spectral index: $\alpha \sim 0.5$ between 1.4 and 5\,GHz and $\alpha\sim 0.8$ between 5 and 8.5\,GHz. If confirmed, this is consistent with optically-thin synchrotron emission, associated with the central AGN.

{\bf J123650+620844}:
This source is classified as a starburst by M05. It does not have a $1.4-8.5$\,GHz spectral index determination. 
A weak unresolved core is detected at 5\,GHz with S/N of 3.5
The peak is slightly offset (0.18$\pm$0.04\,arcsec northwards) with respect
to the 1.4\,GHz peak position, but still overlaps to some extended 
emission detected at the lower frequency.  As with J123624+621128 we have not calculated the 
spectral index. The presence of an offset may support the starburst classification. 
As for J123624+621743, we estimated the probability to have a noise spike within the area corresponding
to the 1.4\,GHz VLA emission above 3.5$\sigma$.
In this case, the probability for a spurious signal is of about 0.3\%.
However, future observations with higher sensitivity and better
{\it uv}-coverage are needed to confirm this detection. 

However, 
as for J123624+621743, the source was not expected to be detected at 5\,GHz:
this source has a flux $S_{(\rm{1.4\,GHz})}\sim$76\,$\mu$Jy and is located well below our 5\,GHz detection threshold (see Fig.~\ref{fig:noise}).
The presence of a significant positional offset between the peaks at
1.4 and 5\,GHz and the apparent flux inconsistency make this detection doubtful.

{\bf J123652+621444}: 
This source is characterized by a flat arcsec-scale spectral index and was classified as an AGN by M05.
The radio structure at 5\,GHz shows a compact core coincident with the peak at 1.4\,GHz. The derived spectral index is inverted
($\alpha\simeq-0.2\pm 0.2$), confirming the AGN nature of the source. 

{\bf J123659+621832}:
This is one of the sources not previously imaged at 1.4\,GHz by M05.
The radio source is identified with a dust-obscured galaxy, an objects showing an IR
excess with respect to the optical emission \citep{Geor11}. 
The object is detected by {\it Chandra} and shows
evidence for moderate absorption. The redshift is not constrained: 
a tentative estimate of a photometric redshift yields a value of 4.4 \citep{Rovilos10}; 
in any case the source must be at redshift $\gsim 2$ \citep{Geor11}.
The arcsec-scale spectral index is flat.
At 5\,GHz we detect a very bright compact source which implies an inverted radio spectrum
between 1.4 and 5\,GHz ($\alpha\sim -0.4$), and a very steep spectrum between 5 and 8.5\,GHz
($\alpha\sim 1.9$). Such spectral behaviour might be due to variability. 
However this kind of spectrum is also typical of Giga-Hertz Peaked radio sources (GPS, e.g. \citealt{Odea}), a class of very compact radio galaxies, thought to be caught in the very early phases of their growth. Whatever the case, this points towards an AGN classification. Assuming a conservative redshift of $z\sim 2$ and a conservative radio spectral index of $\alpha\sim0.26$ (the one  measured between 1.4 and 8.5\,GHz at arcsec scale, see Table~\ref{tab:spix}), we derive a rest-frame frequency $k$-corrected luminosity of $3\times 10^{25}$\,WHz$^{-1}$. Even under such conservative assumptions, this turns out to be the highest radio luminosity source in the sample. 
This potentially very high-redshift source is of particular interest and worth further investigation. 

{\bf J123714+620823}:
The core was detected at 1.4\,GHz by global VLBI observations
\citep{Chi13}, confirming the AGN classification by M05. The arcsec-scale radio spectrum is  flat, indicating 
that the overall radio emission is dominated by the AGN core. This is consistent with our own detection of a bright compact core, with a flat radio spectrum ($\alpha\simeq 0.04\pm 0.09$) between 1.4 and 5\,GHz.

{\bf J123721+620708}:
This is one of the sources not previously imaged at 1.4\,GHz by M05 and does not have a $1.4-8.5$\, spectral index determination.
Our 5\,GHz detection on sub-arcsec scale implies an inverted radio spectrum of $\alpha\sim -0.5$ between 1.4 and 5\,GHz. This may be due to variability, supporting an AGN classification.
However the significance of the 5\,GHz detection is low ($3\sigma$) and deeper observations 
are necessary to confirm the AGN nature of this source.

{\bf J123721+621129}:
The core was detected at 1.4\,GHz by global VLBI observations
\citep{Chi13}, confirming the AGN classification by M05. The arcsec-scale radio spectrum is  inverted, indicating that the overall radio emission is dominated by the AGN core. At 5~GHz we also detect an inverted-spectrum ($\alpha\simeq -0.4\pm 0.2$) unresolved core, that overlaps the peak emission at 1.4\,GHz. 

{\bf J123725+621128}:
This is a wide-angle tailed (WAT) radio galaxy. 1.4\,GHz sub-arcsec scale imaging shows a relatively extended double source. 
At 5\,GHz we detected the eastern hot-spot/jet component of this source. The derived spectral index between 1.4 and 5\,GHz
is steep with $\alpha\simeq 0.90$ and even steeper
between 1.4 and 8.5\,GHz  ($\alpha = 1.36$). The latter,
being derived with a much lower resolution, likely
contains significant contribution from the diffuse tail emission.

\section{Summary \& Future Perspective}
\label{future}

The {\it e-}MERLIN commissioning observations presented here represent the first high resolution wide-field imaging of the GOODS-N field at 5\,GHz. which,
Although limited to 5 antennas of the {\it e-}MERLIN array the observations yield a resolution of 0.2\,arcsec and a depth of about 15\,$\mu$Jy rms noise at the phase centre. 
These observations were used in conjunction with archival data to determine classifications of 15 of 17 sources detected within the field.

We detected 12 out of the 92 sources belonging to the complete sample derived from
deeper 1.4\,GHz MERLIN+VLA observations by M05. Comparing our 5\,GHz images with the MERLIN+VLA images at 1.4\,GHz at the same
resolution we derived for the first time sub-arcsec scale radio spectral indices for 10
of the 12 sources detected. The remaining two sources have displaced peaks at the two frequencies and the 5\,GHz detection needs confirmation.
In addition, we detected 5 more sources located outside the imaged field in M05. 
For such sources our 5\,GHz {\it e-}MERLIN observations provide unique information on the radio source morphology at sub-arcsec scales. 

Due to the still limited sensitivity of the present observations and the lack of
short baselines, our observations are sensitive to relatively bright compact radio
emission, most likely of AGN origin, and indeed many of the detected sources have independent
indicators of AGN activity.
The analysis of 5\,GHz morphologies, together with the derived spectral indices allowed us to 
confirm previous AGN classifications, when available, and propose a classification as AGN/AGN-candidate 
for six of the seven sources previously unclassified. 
Of eight radio sources previously classified as AGN, six have 1.4-5\,GHz spectra flatter than 0.6. 

The fact that most of the detected sources are associated with AGN/AGN-candidates
confirms previous results showing that AGNs dominate at the brighter end of sub-mJy
radio samples (e.g. \citet{Mignano08}).
The relative fraction of detections (and AGN) decreases going to lower flux
densities (AGN sources go from 56$\%$ at $S_{\rm 1.4 GHz}>300\,\mu$Jy, to 
$\ga 16\%$ at $100<S_{\rm 1.4 GHz}<300\,\mu$Jy), where the component associated with
steep-spectrum star-forming galaxies is expected to become more prominent. 

The detected sources span a very wide redshift range ($0.1<z<4.4$). Three of the highest-redshift sources in the detected sample have independent
observational evidence (based on infrared/sub-mm data) that the AGN might be embedded in dusty star-forming galaxies: they are J123642+621331 at $z=4.424$
\citep{Waddi99}, J123649+620737 at $z=2.315$ \citep{Smail04,Casey09} and J123659+621832 at $z>2$ (tentative photometric redshift estimate of $z\sim 4.4$,
\citealt{Rovilos10}). The latter two are among the sources for which sub-arcsec radio imaging was not previously available.  
Interestingly, the radio spectral behaviour of source J123659+621832 seems to mimic that of GPS radio sources. This potentially very high-redshift
 source is of particular interest and is worth further investigation. 
This pilot study provides a indication of what will be achieved in the near future now that {\it e-}MERLIN is fully operational. 
In the following months, the GOODS-N region will be targeted at full resolution (about 50\,mas at 5\,GHz) by deep (sub-$\mu$Jy sensitivity)
 {\it e-}MERLIN 1.4 and 5\,GHz  observations in the framework of the {\it e}MERGE Survey Legacy project \citep{Muxlow08}.
The 5\,GHz observations will provide the resolution to
directly distinguish between jet-like or more compact morphologies,
indicating AGN activity, and less compact radio emission associated with 
star-forming galaxies. 
This analysis will benefit from 1.4\,GHz information, allowing derivation of the source radio 
spectral index on sub-arcsec scales.  Moreover,
ongoing complementary deep Karl G. Jansky Very Large Array (JVLA)
observations will recover the extended
emission unsampled by the {\it e-}MERLIN data on angular scales larger than 0.5\,arcsec.
The combination of deep, high resolution 1.4 and 5\,GHz JVLA and {\it e-}MERLIN observations will provide the opportunity to both spectrally
and morphologically identify embedded low-luminosity AGNs, whilst at the same time map the distribution of star formation activity on scales
from $50-2000$\,mas in the high-redshift galaxy population ($1<z<5$). These angular scales correspond to spatial resolutions spanning
sub-kpc ($\sim 300$\,pc) to tens of kpc at z$>$1.
This strategy will enable us to obtain a bolometrically complete census of star formation, the growth of galaxies and
 their SMBHs, an essential step toward a full understanding of galaxy formation and evolution.

\section*{Acknowledgements}
This work is supported by INAF under grants PRIN-INAF 2009.
{\it e-}MERLIN is a UK National Facility operated by the University of Manchester at Jodrell Bank
Observatory on behalf of PPARC.
DG is grateful to the Jodrell Bank Centre for Astrophysics
and to the Jodrell Bank Observatory for hospitality.
IRS acknowledges support from the Leverhulme Trust and an ERC Advanced Grant Dusrty Gal 321334.

\end{document}